\documentclass{article}
\usepackage{spconf,amsmath,graphicx,hyperref}
\usepackage{amsmath}
\usepackage{amssymb} 
\usepackage{bbm}

\title{Learning Domain-Robust Bioacoustic Representations for Mosquito Species Classification with Contrastive Learning and Distribution Alignment}
\name{Yuanbo Hou$^{1}$, Zhaoyi Liu$^2$, Xin Shen$^1$, Stephen Roberts$^{1}$}
 \address{$^1$University of Oxford, UK  $^2$KU Leuven, Belgium  }
 
\usepackage{hyperref}
\hypersetup{
colorlinks=true,
linkcolor=black
}

\usepackage{caption} 
\usepackage{tabularx}
\usepackage{float}
\usepackage{array}  
\usepackage{booktabs} 
\usepackage{multirow}
\usepackage{makecell}

\usepackage{color}

\usepackage{bbding}

\usepackage{amsfonts,amssymb} 

\usepackage{subfigure} 

\begin{document}
%
\maketitle
\begin{abstract}
Mosquito Species Classification (MSC) is crucial for vector surveillance and disease control. The collection of mosquito bioacoustic data is often limited by mosquito activity seasons and fieldwork. 
Mosquito recordings across regions, habitats, and laboratories often show non-biological variations from the recording environment, which we refer to as domain features.
This study finds that models directly trained on audio recordings with domain features tend to rely on domain information rather than the species'  acoustic cues for identification, resulting in illusory good performance while actually performing poor cross-domain generalization. To this end, we propose a Domain-Robust Bioacoustic Learning (DR-BioL) framework that combines contrastive learning with distribution alignment. Contrastive learning aims to promote cohesion within the same species and mitigate inter-domain discrepancies, and species-conditional distribution alignment further enhances cross-domain species representation. Experiments on a multi-domain mosquito bioacoustic dataset from diverse environments show that the DR-BioL improves the accuracy and robustness of baselines, highlighting its potential for reliable cross-domain MSC in the real world.  
\end{abstract}
\begin{keywords}
Bioacoustics, mosquito species classification, domain shift, contrastive learning
\end{keywords}

\vspace{-2mm}
\section{Introduction}
\label{sec:intro}
\vspace{-2mm}
Bioacoustic Mosquito Species Classification (MSC) aims to identify mosquito species through their flight sound recordings \cite{kiskin2humbugdb}. Different mosquito species transmit distinct pathogens, including malaria, dengue fever, and yellow fever \cite{meerwijk2020phantom}. Accurate MSC is therefore crucial for predicting disease risks and guiding timely interventions.
The advantage of bioacoustic MSC lies in its role as a scalable, non-invasive tool for tracking mosquito populations and understanding their spatio-temporal dynamics \cite{torres1997models}, thereby supporting ecological research.
Compared to traditional methods \cite{gyawali2025morphological} that rely on manual capture and laboratory identification, MSC based on bioacoustics, such as advocated in the HumBug\protect\footnote{\url{https://humbug.ox.ac.uk/}} project \cite{sinka2021humbug}, are more efficient, real-time, and cost-effective \cite{fernandes2021detecting}.

\begin{figure}[t]
    \centering
    \setlength{\abovecaptionskip}{0.1cm}    
	\setlength{\belowcaptionskip}{-0.7cm} 
    \includegraphics [width=0.95\columnwidth]{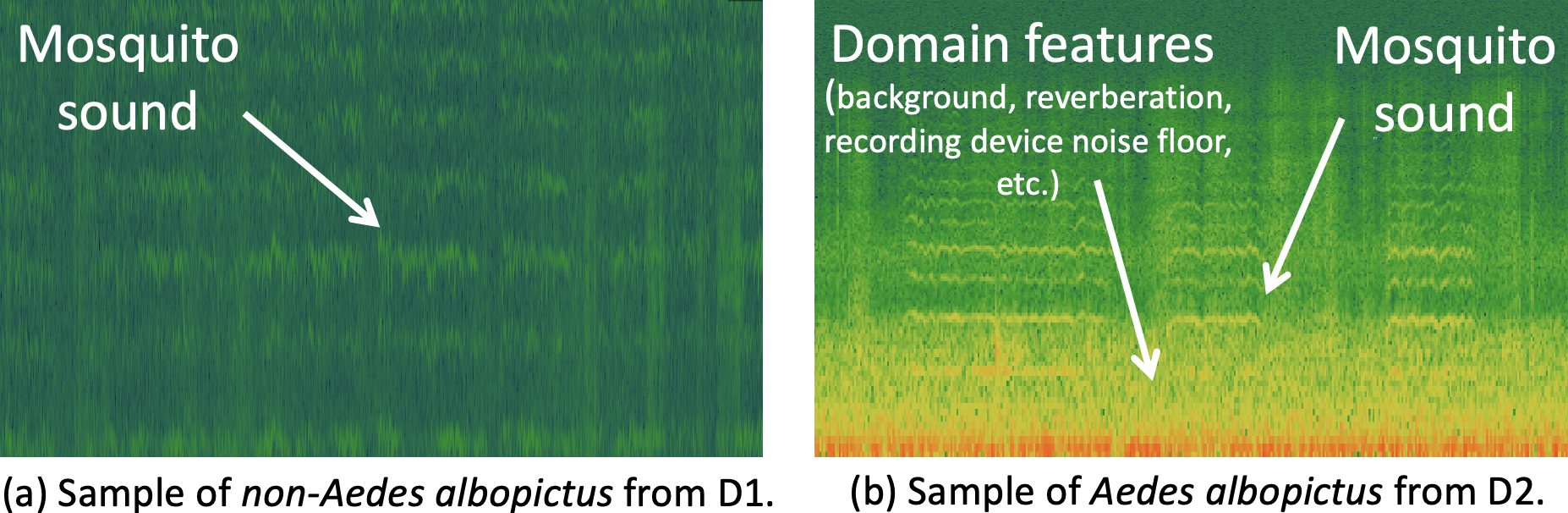}
    \caption{
    Spectrograms from different sources show that the CNN with illusory high test accuracy in Table \ref{tab:cnn_demo} classifies \textit{Aedes albopictus} samples by domain features of D2 rather than species information of \textit{Aedes albopictus}.
    }
    \label{fig_test_demo}
\end{figure}

\begin{figure*}[t]
    \centering
    \setlength{\abovecaptionskip}{0.1cm}    
	\setlength{\belowcaptionskip}{-0.5cm} 
    \includegraphics [width=1.9\columnwidth]{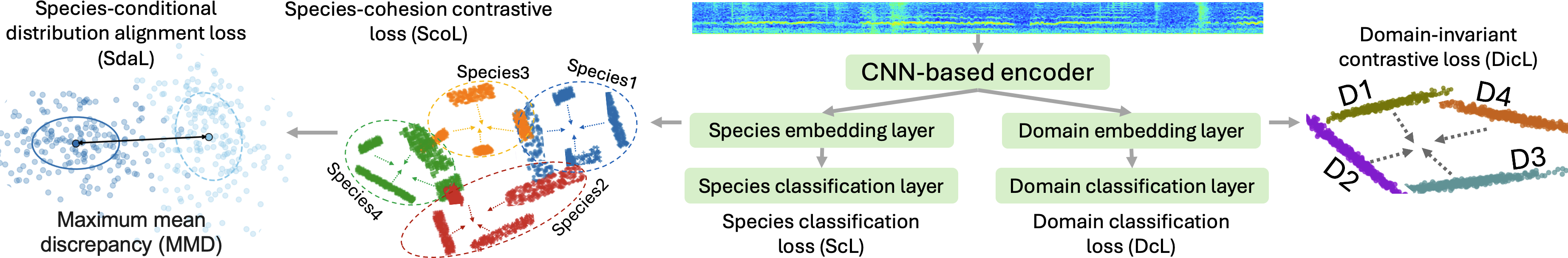}
    \caption{A CNN-based example of an instantiation of the proposed DR-BioL framework.}
    \label{model}
\end{figure*}

Despite the promising prospects of bioacoustic MSC, research in this domain remains challenging. Real bioacoustic data \cite{kiskin2humbugdb}, rather than AI model-generated fake data or artificially synthesized data \cite{11070232}, is scarce, and some species are only active during certain seasons of the year \cite{ loetti2007seasonal}, making data collection time-consuming and laborious.
Furthermore, recordings \cite{kiskin2humbugdb, fernandes2021detecting, mukundarajan2017using, paim2024acoustic} collected across different regions, environments, or laboratories inevitably contain characteristics, such as background noise, recording conditions, or device variations. In this paper, these characteristics are simplified as \emph{domain features}. Models trained directly on these audio files easily overfit to these spurious domain cues rather than learning true bioacoustic information, resulting in illusory high performance and subsequent poor cross-domain generalization.
\begin{table}[b]\footnotesize
	\setlength{\abovecaptionskip}{0cm}   
	\setlength{\belowcaptionskip}{-0.1cm}  
	\renewcommand\tabcolsep{1pt} 
    \vspace{-5mm}
	\centering
	\caption{
    Test accuracy of CNNs on \textit{Ae. albopictus} is compared with and without considering domain features on the training set; details of D1, D2, and D3 are in Section \ref{sec:dataset}.
    }
	\begin{tabular}{   
	p{2.3cm}<{\centering}|
 p{1.8cm}<{\centering}|
	p{1.5cm}<{\centering}|
 p{2.5cm}<{\centering} }
	
		\toprule[1pt] 
		\specialrule{0em}{0.1pt}{0.1pt}  
 
  Training set source & 
  Test set source& 
  CNN & 
  DR-BioL CNN   \\ 
   
\hline 
	  D1 + D2 & D1 + D2 &  99.79 \% & 92.81 \%  \\  
      D1 + D2 & D3 &  41.40 \% & 74.92 \%  \\  
		\specialrule{0em}{0pt}{0em}
		\bottomrule[1pt]
	\end{tabular}
	\label{tab:cnn_demo}
\end{table}   
In Table \ref{tab:cnn_demo}, Domain 1 (D1) contains data for 7 species, while data for another species, \textit{Aedes (Ae) albopictus} \cite{loh2024differences}, comes from Domain 2 (D2).
A Convolutional Neural Network (CNN) \cite{he2016deep} shows high accuracy for \textit{Ae albopictus} on a test set consisting of D1 and D2 data.
However, for \textit{Ae albopictus} data from the D3 (new domain), the CNN's recognition accuracy dropped significantly to 41.40\%.
This drop in model performance stems from the difference in data distribution between the source and target domains, which is known as domain shift \cite{zhang2021adaptive}.
In contrast, under the same conditions, the domain-aware CNN, proposed in this paper, performs considerably better.
Fig.~\ref{fig_test_demo}  illustrates that, as bioacoustics data contain domain features indicative of their source, these features can mislead model learning and hinder generalization ability, thereby reducing the reliability of such classification models in real-world applications.


  
To address these challenges, we propose domain-robust bioacoustic learning (DR-BioL), a framework for MSC using bioacoustic data collected from diverse sources. 
In contrast to Domain Adversarial Training (DAT) methods \cite{ganin2016domain} for domain shift \cite{zhang2021adaptive}, DR-BioL employs contrastive learning \cite{khosla2020supervised} to enhance cross-domain species consistency by promoting cohesion among representations of the same species across different domains, while simultaneously maximizing the separation between different species.
Furthermore, the species-conditional distribution alignment is incorporated to stabilize species-level representations across domains.

The contributions are as follows:
1) We propose DR-BioL, a framework that integrates species discrimination and domain robustness for MSC.
2) To enforce species discriminability while promoting domain invariance, we introduce a contrastive learning-based species cohesion loss, consisting of species-discriminative and domain-invariant losses. To align species-level distributions across domains, a conditional distribution alignment loss is introduced.
3) We validate DR-BioL on multi-domain bioacoustic datasets.

\vspace{-2mm}
\section{Domain-Robust Bioacoustic Learning}
\label{sec:Method}

\vspace{-2mm}
The proposed Domain-Robust Bioacoustic Learning (DR-BioL) framework consists of a bioacoustic encoder and five complementary optimization objectives, as shown in Fig. \ref{model}.

\vspace{-4mm}
\subsection{Bioacoustic representation encoder}\label{sec:encoder}
\vspace{-1mm}
Given the excellent performance of CNN-based models in previous MSC studies \cite{kiskin2humbugdb, paim2024acoustic, fernandes2021detecting}, the DR-BioL instantiation in Fig. \ref{model} uses a CNN as the bioacoustic representation encoder. The CNN-based encoder consists of 4 layers of VGG-like \cite{vgg} convolutional blocks with 64, 128, 256, and 512 filters, respectively. Each convolutional block contains 2 convolutional layers with a kernel size of (3 × 3). Batch normalization \cite{batchnormal} and ReLU activation functions \cite{relu} are used to accelerate and stabilize the training.

\vspace{-4mm}
\subsection{Species classification loss (ScL)}\label{sec:Species_classification}   
\vspace{-1mm}
Following the encoder, a species embedding layer and a species classification layer, each consisting of a Fully Connected (FC) layer with 512 and $N_S$ units, respectively, are used to learn target-oriented representations and perform the MSC task. $N_S$ is the number of mosquito species. 
Since multiple mosquito species may occur simultaneously in real-world scenarios, a sum of binary cross-entropy losses \cite{kong2020panns} is used as the mosquito Species classification Loss (ScL) between the species prediction $\hat{y}_{s}\in\mathbb{R}^{N_S}$ from the last layer and the label ${y}_{s}\in\mathbb{R}^{N_S}$. 
\begin{equation}
\label{eq:ScL}
\setlength{\abovedisplayskip}{1pt}
\setlength{\belowdisplayskip}{1pt}
\mathcal{L}_\mathrm{ScL} = - \sum\nolimits_{i=1}^{N_S} 
{y}_{s_i}\log(\hat{y}_{s_i}) + (1-{y}_{s_i})\log(1-\hat{y}_{s_i})
\end{equation}
 
\vspace{-4mm}
\subsection{Domain classification Loss (DcL)} 
\vspace{-2mm}
Similarly, the Domain Classification (DC) branch consists of domain embeddings and classification layers based on FC layers, containing 256 and $N_D$ units, respectively. $N_D$ is the number of domains.
Since each audio clip has a unique source, the DC is a single-label multi-class classification task. The cross entropy loss \cite{hou2023cooperative} is used as the domain classification loss (DcL) between the domain prediction $\hat{y}_{d}\in\mathbb{R}^{N_D}$ and the label ${y}_{d}\in\mathbb{R}^{N_D}$.
\begin{equation}
\setlength{\abovedisplayskip}{1pt}
\setlength{\belowdisplayskip}{1pt}
\mathcal{L}_\mathrm{DcL} = - \sum\nolimits_{j=1}^{N_D} 
{y}_{d_j}\log(\hat{y}_{d_j})
\end{equation}

\vspace{-4mm}
\subsection{Contrastive cross-domain species cohesion loss} 
\vspace{-2mm}
Cross-domain species cohesion loss relies on supervised contrastive learning \cite{khosla2020supervised} from both species and domain perspectives, prompting the acoustic encoder to mitigate interference from domain features and learn robust cross-domain species representations.
For the anchor $i$ in the supervised contrastive learning \cite{khosla2020supervised}, define $A(i)=\{1,\dots,N\}\setminus\{i\}$. $A(i)$ does not contain the anchor $i$.
Given $z_{\{i, p, a\}}$ are embeddings, the supervised-contrastive per-anchor objective with a chosen positive index set $P(i)\subseteq A(i)$ is
\begin{equation}
\setlength{\abovedisplayskip}{1pt}
\setlength{\belowdisplayskip}{1pt}
\label{eq:sup}
\begin{aligned}
\mathcal{L}^{\mathrm{sup}}_i\big(P(i)\big)
&= - \log \sum_{p \in P(i)} \exp\!\big(\mathrm{sim}(z_i, z_p)/\tau\big) \\
&\quad + \log \sum_{a \in A(i)} \exp\!\big(\mathrm{sim}(z_i, z_a)/\tau\big)
\end{aligned}
\end{equation}
where $\mathrm{sim}(u,v)=u^\top v$, $\tau$ is the temperature term in contrastive learning \cite{khosla2020supervised}, $\tau$ defaults to 0.01.
We combine two instantiations of this objective to obtain a representation that is species-cohesive and domain-robust.

\textbf{Species-cohesion contrastive Loss (ScoL):}
To enforce intra-class compactness and inter-class separation, we set as positive all samples sharing the class label with the anchor:
\begin{equation}
\setlength{\abovedisplayskip}{1pt}
\setlength{\belowdisplayskip}{1pt}
\label{eq:Pclass}
P_{\mathrm{species}}(i) \;=\; \big\{\, p \in A(i)\;\big|\; y_p^{\mathrm{species}} = y_i^{\mathrm{species}} \,\big\}
\end{equation}
The ScoL averages the per-anchor objectives over anchors with at least one positive, $\mathcal{I}_{\mathrm{species}}=\{i \mid |P_{\mathrm{species}}(i)|>0\}$,
\begin{equation}
\setlength{\abovedisplayskip}{1pt}
\setlength{\belowdisplayskip}{1pt}
\label{eq:Lclass}
\begin{aligned}
\mathcal{L}_{\mathrm{ScoL}}
&=\frac{1}{|\mathcal{I}_{\mathrm{species}}|}\sum_{i \in \mathcal{I}_{\mathrm{species}}}
\mathcal{L}^{\mathrm{sup}}_i\!\big(P_{\mathrm{species}}(i)\big)
\end{aligned}
\end{equation}

\textbf{Domain-invariant contrastive Loss (DicL):}
To suppress domain-specific variability, we set as positive all samples drawn from \text{different} domains than the anchor:
\begin{equation}
\setlength{\abovedisplayskip}{1pt}
\setlength{\belowdisplayskip}{1pt}
\label{eq:Pdomain}
P_{\mathrm{domain}}(i) \;=\; \big\{\, p \in A(i)\;\big|\; y_p^{\mathrm{domain}} \neq y_i^{\mathrm{domain}} \,\big\}.
\end{equation}
Then, Given $\mathcal{I}_{\mathrm{domain}}=\{i \mid |P_{\mathrm{domain}}(i)|>0\}$, the DicL is
\begin{equation}
\setlength{\abovedisplayskip}{1pt}
\setlength{\belowdisplayskip}{1pt}
\label{eq:Ldomain}
\begin{aligned}
\mathcal{L}_{\mathrm{DicL}}
&=\frac{1}{|\mathcal{I}_{\mathrm{domain}}|}\sum_{i \in \mathcal{I}_{\mathrm{domain}}}
\mathcal{L}^{\mathrm{sup}}_i\!\big(P_{\mathrm{domain}}(i)\big)
\end{aligned}
\end{equation}

\vspace{-3mm}
\subsection{Species-conditional distribution alignment loss} 
\vspace{-1mm}
The Species-conditional distribution alignment Loss (SdaL) builds upon the representations learned by ScoL and aims to explicitly align their distributions for each species.
SdaL employs the Maximum Mean Discrepancy (MMD) \cite{tolstikhin2016minimax} metric, minimizing MMD to bring the distributions of representations within the same species closer together, thereby learning robust species representations across domains.
Given that $S_{c_{i1}}$ and $S_{c_{i2}}$ are embeddings of sample \{1, 2\} from the same species $C_i$, $C_i$ is one of the classes in $N_S$ in Eq. (\ref{eq:ScL}), SdaL is defined as
\begin{equation}
\setlength{\abovedisplayskip}{1pt}
\setlength{\belowdisplayskip}{1pt} 
\mathcal{L}_{\mathrm{SdaL}}=\frac{1}{N_S}\sum_{C_n \in N_s}\frac{1}{|C_n|}\sum_{c_i \in Cn}{\mathrm{MMD}^2(S_{c_{i1}}, S_{c_{i2}})},
\end{equation}
where $\mathrm{MMD}^2(a, b)=k_\sigma(a, a) + k_\sigma(b, b) - 2k_\sigma(a, b)$, and $k_\sigma()$ defaults to the Radial Basis Function (RBF) kernel \cite{larsson2024scaling}.

\vspace{-2mm}
\subsection{Total loss} 
\vspace{-1mm}
The final loss function of DR-BioL is given by the weighted sum of the separate loss functions:
\begin{equation}
\setlength{\abovedisplayskip}{1pt}
\setlength{\belowdisplayskip}{1pt}
\mathcal{L} = \lambda_1\mathcal{L}_\mathrm{ScL} + \lambda_2\mathcal{L}_\mathrm{ScoL}  +\lambda_3\mathcal{L}_\mathrm{SdaL} + \lambda_4\mathcal{L}_\mathrm{DcL} + \lambda_5\mathcal{L}_\mathrm{DicL},
\end{equation}
where $\lambda_i$ is the scale factor of each loss, $\lambda_i$ defaults to 1. \{$\lambda_1, \lambda_2, \lambda_3$\} aim to optimize species-related representations, \{$\lambda_4, \lambda_5$\} focus on domain-related representations.
Various configurations of $\lambda_i$ are explored in the experiments below.

\vspace{-2mm}
\section{Experiments and results}
 
\vspace{-2mm}
\subsection{Dataset, experiments setup, and metrics}\label{sec:dataset}
\begin{figure}[t]
    \centering
    \setlength{\abovecaptionskip}{0.05cm}    
	\setlength{\belowcaptionskip}{-0.5cm} 
    \includegraphics [width=1\columnwidth]{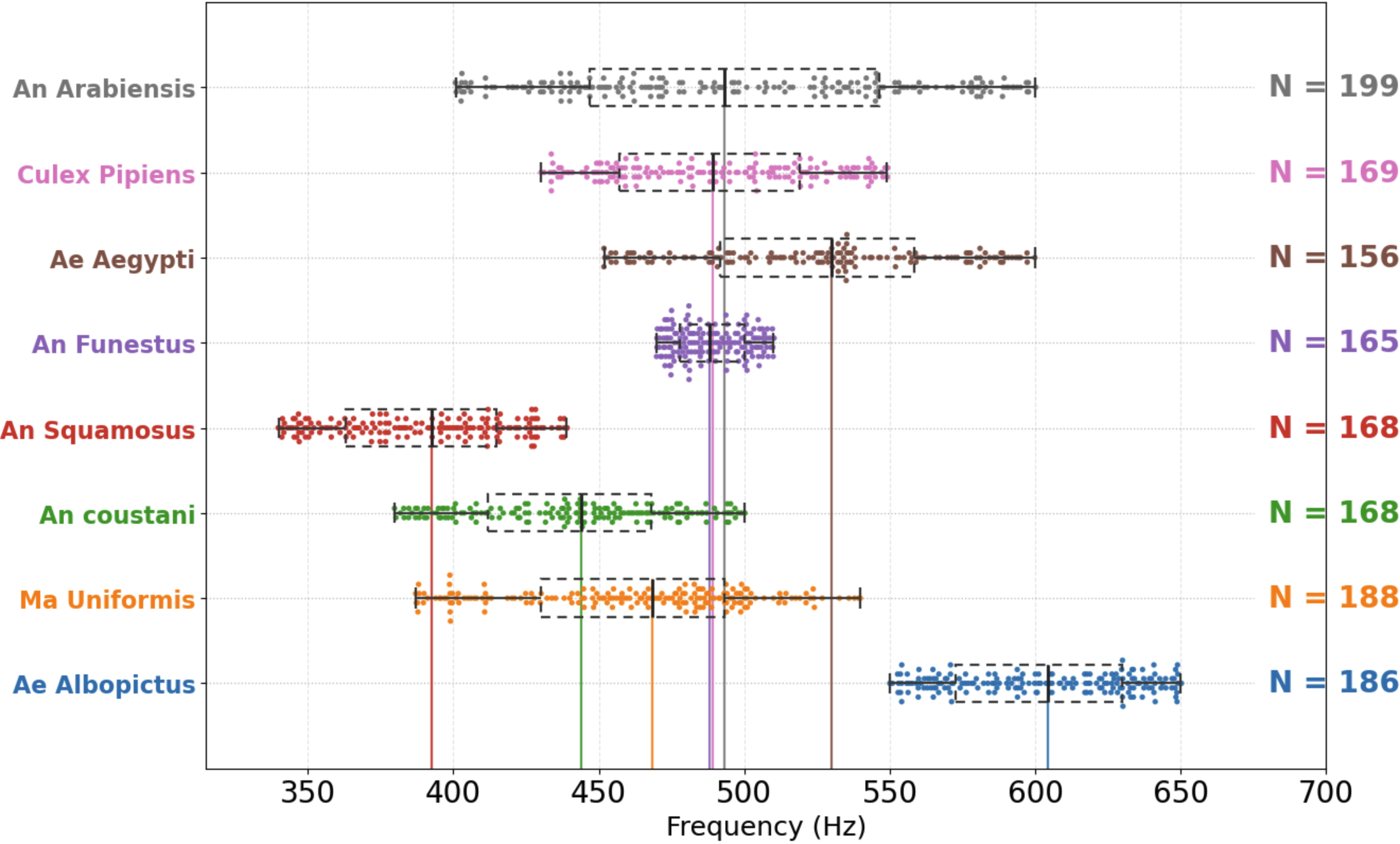}
    \caption{Distribution of wingbeat frequencies for the mosquito species data used in this paper.}
    \label{species_f}
\end{figure}

\vspace{-2mm}
We utilize mosquito audio datasets recorded in four different countries and multiple regions to create a multi-domain mosquito dataset comprising of eight species: \textit{An Arabiensis}, \textit{Culex Pipiens}, \textit{Ae Aegypti}, \textit{An Funestus}, \textit{An Squamosus}, \textit{An coustani}, \textit{Ma Uniformis}, \textit{Ma Africanus},  \textit{Ae Albopictus}. 
The first seven species are from the HumBug dataset \cite{kiskin2humbugdb} recorded in Tanzania, denoted as Domain 1 (D1), comprising 37688 audio clips, totalling about 20.94 hours.  
The Kasetsart dataset (denoted D2) of \textit{Ae Albopictus} recorded in Thailand, 
which is part of the HumBug project \cite{sinka2021humbug},
contains 655 audio clips, totalling about 0.37 hours.
The UFRGS dataset \cite{paim2024acoustic} (denoted as D3) of male and female mosquito \textit{Ae Aegypti} and \textit{Ae Albopictus} recorded in Brazil contains 16727 audio clips with a total duration of about 9.30 hours. 
The Abuzz dataset \cite{mukundarajan2017using} (denoted as D4) of mosquito \textit{An Arabiensis}, \textit{Culex Pipiens}, \textit{Ae Aegypti}, and \textit{Ae Albopictus} recorded in the USA contains 5054 audio clips with a total duration of about 2.81 hours.
The total duration of the four-domain MSC dataset used in this paper is 33.42 hours.
Fig. \ref{species_f} shows the distribution of wingbeat frequencies for the 8 mosquito species used in this paper, from audio clips randomly selected from the dataset.
In our experiments, the duration of training, validation, and test sets is 23.46, 4.26, and 5.70 hours, respectively. 

The acoustic features are 64-bank log-mel energies \cite{kong2020panns}, extracted with a 64 ms Hamming window and 10 ms overlap. Training uses batch size 64 and AdamW \cite{adamw} with learning rate 0.0005.
Dropout, normalization, and early stopping \cite{dropout} are applied to prevent overfitting; training stops if validation MSC accuracy does not improve within 10 epochs after the 50th, with a maximum of 500 epochs. Each model is trained 10 times to report the mean performance. MSC is evaluated by Accuracy (Acc.), Average Precision (AP) \cite{lipton2014optimal}, and AUC \cite{metrics}.
For source dataset details, code, and models, please visit the \textbf{\textit{homepage}}
{\footnotesize{(\textcolor{blue}{\underline{https://github.com/Yuanbo2020/DR-BioL}})}}.
    
\vspace{-4mm}
\subsection{Results and analysis}

\begin{table}[b]\footnotesize 
	\setlength{\abovecaptionskip}{0cm}   
	\setlength{\belowcaptionskip}{-0.2cm}  
	\renewcommand\tabcolsep{1pt} 
    \vspace{-3mm}
	\centering
	\caption{Ablation study of DR-BioL on the validation set.}
	\begin{tabular}{   
	p{0.5cm}<{\centering}|
    p{0.8cm}<{\centering}|
    p{0.8cm}<{\centering}|
    p{0.8cm}<{\centering}|
    p{0.8cm}<{\centering}|
    p{0.8cm}<{\centering}|
    p{1.8cm}<{\centering}| 
    p{1.7cm}<{\centering}}
	
		\toprule[1pt] 
		\specialrule{0em}{0.1pt}{0.1pt}  

 \multirow{2}{*}{\makecell[c]{\#}} &
 \multicolumn{3}{c|}{Mosquito species} & 
 \multicolumn{2}{c|}{Domain} &  
 \multirow{2}{*}{\makecell[c]{Acc.} (\%)} & 
 \multirow{2}{*}{\makecell[c]{AP}}  \\ 

\cline{2-6} 
  & $\mathcal{L}_\mathrm{ScL}$ & $\mathcal{L}_\mathrm{ScoL}$ & $\mathcal{L}_\mathrm{SdaL}$ &  $\mathcal{L}_\mathrm{DcL}$  & $\mathcal{L}_\mathrm{DicL}$ & & \\ 
\hline 
	  1 & \CheckmarkBold & \CheckmarkBold & \CheckmarkBold & \CheckmarkBold & \CheckmarkBold & 82.189 $\pm$ 0.215 & 0.884 $\pm$ 0.001  \\  
       2 & \CheckmarkBold & \XSolidBrush & \CheckmarkBold & \CheckmarkBold & \CheckmarkBold & 81.253 $\pm$ 0.639  & 0.881 $\pm$ 0.004  \\  
    3 & \CheckmarkBold & \XSolidBrush & \XSolidBrush & \CheckmarkBold & \CheckmarkBold & 80.571 $\pm$ 0.453 &  0.873 $\pm$ 0.003  \\  
    4 & \CheckmarkBold & \CheckmarkBold & \CheckmarkBold & \XSolidBrush & \CheckmarkBold & 81.731 $\pm$ 0.372 &   0.883 $\pm$ 0.006  \\  
    5 & \CheckmarkBold & \CheckmarkBold & \CheckmarkBold & \XSolidBrush & \XSolidBrush & 82.683 $\pm$ 1.183 &   0.887 $\pm$ 0.013  \\  
		\specialrule{0em}{0pt}{0em}
		\bottomrule[1pt]
	\end{tabular}
	\label{tab:ablation}
\end{table}

\vspace{-2mm}
\noindent
\textbf{Ablation study.}  
DR-BioL contains five losses, so the first experiment explores \textit{which of these five losses has the greater impact on MSC performance}.
Using \#1 in Table \ref{tab:ablation} as the baseline reference, the model's accuracy on the MSC task progressively declines as mosquito species-related losses (\#2 and \#3) are removed. 
Conversely, removing domain-feature-related losses in \#5 led to improved accuracy on the MSC task. 
Similar to the results in Table \ref{tab:cnn_demo}, the improvement in \#5 stems not from reliance on mosquito species information, but from leveraging domain features implicitly embedded within audio samples from different sources. 
The improvement in \#5 demonstrates that without the constraints of the domain-related losses, the model can easily rely on relatively easier-to-distinguish domain features for mosquito classification.
In short, the results in Table \ref{tab:ablation} indicate that mosquito species-related loss is more important in the MSC task. 
To achieve robust cross-domain MSC, the model must strike a balance between mosquito species-related loss and domain-related loss.

\vspace{-1mm}
\begin{table}[H]\footnotesize 
	\setlength{\abovecaptionskip}{0cm}   
	\setlength{\belowcaptionskip}{-0.2cm}  
	\renewcommand\tabcolsep{1pt} 
	\centering
	\caption{Effect of different $\lambda_i$ values on the validation set.}
	\begin{tabular}{   
	p{0.5cm}<{\centering}|
    p{0.8cm}<{\centering}|
    p{0.8cm}<{\centering}|
    p{0.8cm}<{\centering}|
    p{0.8cm}<{\centering}|
    p{0.8cm}<{\centering}|
    p{1.8cm}<{\centering}| 
    p{1.7cm}<{\centering}}
	
		\toprule[1pt] 
		\specialrule{0em}{0.1pt}{0.1pt}  

 \multirow{2}{*}{\makecell[c]{\#}} &
 \multicolumn{3}{c|}{Mosquito species} & 
 \multicolumn{2}{c|}{Domain} &  
 \multirow{2}{*}{\makecell[c]{Acc.} (\%)} & 
 \multirow{2}{*}{\makecell[c]{AP}}  \\ 

\cline{2-6} 
  & $\lambda_1$ & $\lambda_2$ & $\lambda_3$ &  $\lambda_4$  & $\lambda_5$ & & \\ 
\hline 
	  1 & 1 & 1 & 1 & 1 & 1 & 82.189 $\pm$ 0.215 & 0.884 $\pm$ 0.001  \\  
      2 & 1 & 1 & 1 & 0.01 & 1 & 83.902 $\pm$ 0.302 & 0.891 $\pm$ 0.006  \\  
      3 & 1 & 1 & 1 & 0.01 & 0.1 & \textbf{84.644} $\pm$ 0.305 & \textbf{0.904} $\pm$ 0.007  \\  
      4 & 1 & 0.1 & 1 & 0.01 & 0.1 & 84.271 $\pm$ 0.342 & 0.893 $\pm$ 0.009  \\  
      5 & 1 & 0.1 & 1 & 0.1 & 0.1 & 83.975 $\pm$ 0.194 & 0.896 $\pm$ 0.008  \\  
      6 & 1 & 0.1 & 0.1 & 0.1 & 0.1 & 84.135 $\pm$ 0.434 & 0.887 $\pm$ 0.002  \\  
      
		\specialrule{0em}{0pt}{0em}
		\bottomrule[1pt]
	\end{tabular}
	\label{tab:alpha}
\end{table}

\vspace{-4mm}
\noindent
\textbf{Performance of different combinations of weights $\lambda_i$.}
Optimizing the losses in DR-BioL is challenging because different metrics are calculated differently, and each loss acts on different components of the model.
Table \ref{tab:alpha} intuitively presents the results of different weight combinations by adjusting parameters empirically.
\#2 reduces the weight of $\lambda_4$ for $\mathcal{L}_\mathrm{DcL}$, thereby reducing the constraint for the model to learn domain-specific features to discriminate domains. 
This reduces the model's focus on domain features and achieves better results than \#1. 
\#3 further reduces the weight of $\lambda_5$ for $\mathcal{L}_\mathrm{DicL}$, thereby lowering the constraint for the model's focus on fusing representations from different domains of the same species. 
This allows the bioacoustic encoder to strike a balance between learning cross-domain species-cohesion representations and learning domain-invariant representations. 
For the rest, reducing the mosquito species-related weights will decrease the model's MSC performance.  
DR-BioL is required to prioritize the weights for species-related representations while also paying sufficient attention to domain features to learn domain-robust bioacoustic representations.

\vspace{-1mm}
\begin{table}[H] \footnotesize 
\setlength{\abovecaptionskip}{0.0cm}   
	\setlength{\belowcaptionskip}{-0.2cm}
	\renewcommand\tabcolsep{1pt} 
	\centering
	\caption{Comparison of MSC results on the test set.}
	\begin{tabular}{
	p{0.3cm}<{\centering}| 
 p{1.7cm}<{\centering}| 
 p{1.5cm}<{\centering} |
	p{1.5cm}<{\centering}| 
 p{1cm}<{\centering}|
 p{1cm}<{\centering}|
 p{1cm}<{\centering}
	} 
	   \toprule[1pt] 
    \specialrule{0em}{0.1pt}{0.1pt} 

  \multirow{1}{*}{\makecell[c]{\#}} & \multirow{1}{*}{\makecell[c]{Model}} &  Param.(M) &  
FLOPs (G) & \multirow{1}{*}{\makecell[c]{Acc.} (\%)} & \multirow{1}{*}{\makecell[c]{AUC}} & \multirow{1}{*}{\makecell[c]{AP}} \\

  \hline 

   1 & Baseline CNN  & 4.9530 & 2.6152     &  80.031 & 0.9680 & 0.8616  \\
    
2 & CNN-Trans.  & 1.5606 & 0.0824    &   74.327	 & 0.9569	 & 0.8316     \\

3 & YAMNet  & 3.2147 & 0.0052     & 77.360 & 	0.9591	 & 0.8332     \\

4 & MobileNetV2  & 2.2335 & 0.0738     & 76.307 & 	0.9543 & 	0.8206   \\

5 & PANNs & 79.6902 &  3.9787  & 81.679	 & 0.9653	 & 0.8511 \\
 
6 & DAT CNN  & 5.0854 & 2.6155   &   79.583	 & 0.9607 & 	0.8481\\
\hline 
 7 & DR-BioL & 5.0854 &  2.6155 &  \textbf{85.345}	 & \textbf{0.9732} & \textbf{0.9002}	\\
	 
	\specialrule{0em}{0pt}{0em}
		\bottomrule[1pt]
	\end{tabular}
	\label{tab:other_models}
\end{table}

\vspace{-4mm}
\noindent
\textbf{Comparison to other methods.}
Table \ref{tab:other_models} shows comparative results from several different models on the multi-domain multi-species mosquito dataset, including CNNs that performed well in previous MSC-related studies \cite{kiskin2humbugdb, fernandes2021detecting, mukundarajan2017using, paim2024acoustic}.
The baseline CNN consists directly of the bioacoustic encoder from Section \ref{sec:encoder}, plus the mosquito species classification branch from Section \ref{sec:Species_classification}. 
\#2 adopts the CNN-plus-Transformer architecture that performs well in bioacoustic tasks \cite{fundel2023automatic}, adding a Transformer encoder between the bioacoustic encoder and the species classification layer.
YAMNet and MobileNetV2 \cite{sandler2018mobilenetv2} are classic and efficient CNN classification models. 
Leveraging weights trained on the large-scale 5800-hour dataset AudioSet, PANNs \cite{kong2020panns} achieve excellent performance on diverse audio-related tasks.
It is noteworthy that DAT \cite{ganin2016domain} CNN, like DR-BioL, equally aims to learn cross-domain species representations. As \#6 and \#7 are based on the same species-domain dual-branch CNN model with differing losses, their parameter (Param.) counts and computational load (FLOPs) are identical. However, DR-BioL, which employs contrastive learning and distribution alignment, achieves superior results on the test set.

\begin{figure}[t]
    \centering
    \setlength{\abovecaptionskip}{0.05cm}    
	\setlength{\belowcaptionskip}{-0.6cm} 
    \includegraphics [width=1\columnwidth]{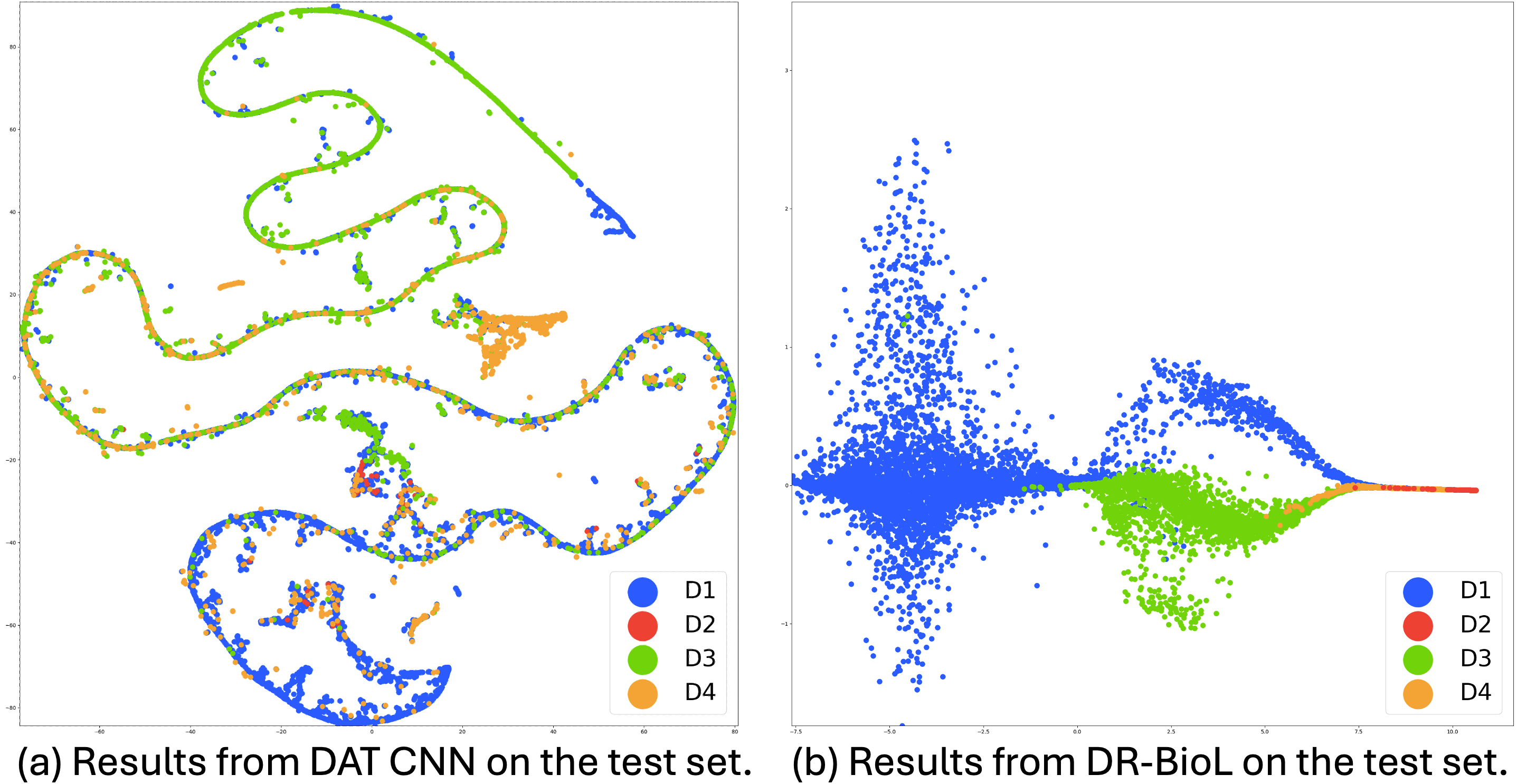}
    \caption{Visualization of the domain embeddings using t-SNE.}
    \label{domain_emb}
\end{figure}

\noindent
\textbf{Discussion.} 
DAT \cite{ganin2016domain} is a typical approach to addressing domain shift \cite{zhang2021adaptive}. To intuitively compare DAT with DR-BioL, Fig. \ref{domain_emb} visualizes their domain embeddings. 
In (a), embeddings of DAT CNN are obfuscated due to the forced effect of the gradient reversal layer in DAT \cite{ganin2016domain}, resulting in a compression of domain information and a degeneration of the distribution into a mixed curve. 
While embeddings of DR-BioL in (b) tend to converge across domains under the contrastive learning constraint, embeddings of D2 and D4 converge and extend to the line connecting D1 and D3, some inter-domain structure is still preserved.
DAT achieves domain invariance by strictly suppressing domain features, and this indiscriminate obfuscation can also erase some species-related cues. 
As a result, the bioacoustic representation, while domain-inseparable, is limited in expressiveness, weakening species separability. This explains the slightly lower performance of DAT CNN compared to Baseline CNN in Table \ref{tab:other_models}.

The contrastive learning constraint in DR-BioL is more flexible. 
Rather than forcibly eliminating all domain-related variation, DR-BioL guides the model to prioritize species discrimination cues and mitigate the impact of domain differences. 
This balance enables the model to capture fine-grained acoustic features for MSC while reducing reliance on domain features. As a result, DR-BioL shows better results in both cross-domain robustness and species classification accuracy.

\vspace{-4mm}
\section{CONCLUSION}
\label{sec:CONCLUSION} 
\vspace{-2mm}
 
We present DR-BioL, a framework that integrates species discrimination with domain robustness for MSC. 
By uniting contrastive species cohesion, species-conditional alignment, and domain-invariant contrasts, it achieves a balance between discriminability and cross-domain invariance. Experiments demonstrate superior performance over baselines and DAT, preserving species cues while mitigating domain dependence, underscoring its potential for reliable bioacoustic monitoring.


\section{ACKNOWLEDGEMENTS}
\label{sec:ACKNOWLEDGEMENTS}
The authors thank the pan-continent HumBug$^1$ team for the extensive field collection and curation of the HumBugDB database of multiple mosquito species. Yuanbo Hou and Stephen Roberts are grateful for funding from the UK Natural Environment Research Council, Grant APP17496.


\bibliographystyle{IEEEbib}
\bibliography{refs}

\end{document}